\documentclass[prl,superscriptaddress,showkeys,showpacs,twocolumn]{revtex4-1}
\usepackage{graphicx}
\usepackage{latexsym}
\begin {document}
\title {Phase transition and fast agreement in Naming Game with preference for multi-word agents}
\author{Dorota Lipowska}
\affiliation{Faculty of Modern Languages and Literature, Adam Mickiewicz University, Pozna\'{n}, Poland}
\author{Adam Lipowski}
\affiliation{Faculty of Physics, Adam Mickiewicz University, Pozna\'{n}, Poland}
\begin {abstract}
We examine a variant of the Naming Game, where agents having several words communicate more often than single-word agents. Depending on the preference and dimensionality, the model either converges to a single-language  state as in an ordinary Naming Game or  remains in a disordered, multi-language phase. At the transition point separating these regimes, due to a percolation-like process, the model converges to a single-language state but much faster than in the ordinary naming game. We also show that the coarsening dynamics of the ordinary Naming Game is slower than expected due to stripe structures that sometimes spontaneously form during the evolution of the model.  
\end{abstract}

\maketitle
\section{Introduction}
Recently, statistical mechanics methods are frequently used to describe phenomena outside the realm of traditional physics \cite{castellano}. Indeed, models similar to those used to analyse equilibrium or nonequilibrium  properties  of condensed matter, liquids, or gases are now successfully applied to describe opinion formation, crowd behaviour or culture dynamics.  This is not unexpected -- similarly to particle systems, societies or populations might be also considered as composed of many interacting subunits (people, organisms), thus forming complex dynamical systems, where statistical-mechanics phenomenology proved to be so useful. Since language might be considered as an emergent property of a system of communicating (interacting) agents, statistical-mechanics modeling finds interesting applications also  in linguistics, allowing us to analyse processes like language emergence, its evolution or extinction \cite{schulze}.   

One of the most basic linguistic processes that attracts considerable interdisciplinary efforts \cite{nowak} is the agreement dynamics. Due to this process, a population of locally interacting agents can establish, without any global control, commonly shared conventions. An important model of the agreement dynamics was introduced by Steels and is known as the Naming Game~\cite{steels}. Subsequently, Baronchelli {\it et al.} \cite{BARONCHELLI2006} introduced its simplified but frequently analysed version, the so-called minimal Naming Game. In Naming Game, a population of agents take part in successive communication attempts, which eventually leads to the emergence of a common vocabulary or even more complex forms like categories \cite{puglisi}, or grammar \cite{beuls}. The Naming Game is supposed to mimic processes responsible for the emergence of language \cite{steels2011,liplip2007} but it was also used to describe opinion dynamics of large-scale autonomously operating
wireless sensor networks \cite{lu}, leader election mechanisms \cite{baron2011}, or formation of a tagging scheme in a group of robots \cite{steels1997}. Due to numerous studies on the Naming Game \cite{nolfi}, it became a paradigmatic model of semiotic dynamics and acquired status similar perhaps to that of the Prisoner Dilemma in agent-based socio-economic studies \cite{axelrod}. 

A primary point of interest in Naming Game studies is the time needed to reach an agreement. As expected, it depends on the network of interactions between agents. Using analogies with a random walk and a coarsening dynamics of other nonconservative systems~\cite{bray}, Baronchelli {\it et al.} \cite{baron2006} predicted that for $d$-dimensional ($d\leq 4$) lattices of linear size~$L$  the characteristic time $\tau$ (normalized per number of agents $N=L^d$) needed to reach an agreement in the entire system should scale as
\begin{equation}
\vspace{-3mm}
\tau \sim L^{\alpha},  \ \ \alpha=2.
\label{barontau}
\end{equation}
For the Naming Game on a complete graph (of infinite dimension),  the predicted scaling~\cite{baron2006} $\tau \sim N^{\frac{1}{2}}$ is equivalent to Eq.~(\ref{barontau}) with $d=4$. A behaviour similar to the case of a complete graph was found for complex networks with small-world \cite{dallasta} or scale-free topologies \cite{scalefree}. 

Equation (\ref{barontau}) is basically an inversion of the expression for the diffusively growing length scale $l\sim t^{1/2}$ and the requirement that $l$ reached the system size $L$. In our opinion, however,  such a reasoning should be taken with care. Indeed, even in the simplest Ising model, the low-temperature coarsening dynamics is known to lead (sometimes) to  stripe-like (zero-curvature) structures that slow down or even trap the dynamics~\cite{lip1999,olejarz-prl}. Formation of such structures severly affects the size-dependence of quantities like $\tau$ and, at least in the Ising model, invalidates Eq.~(\ref{barontau}). 

Taking into  account  a possible formation of stripe structures and the importance of a Naming Game, it would be interesting to reexamine whether Eq.~(\ref{barontau})  describes properly the size-dependence of the characteristic time~$\tau$. Potential applications suggest that it would also be desirable to search for other variants of a Naming Game with possibly faster agreement dynamics. Let us notice that the Naming Game is a model with an absorbing state (once the system reaches an agreement, it stays in such a state forever). Models of this sort are known to have a rich dynamics mainly due to  phase transitions between active and absorbing phases~\cite{haye}. 
As a potential candidate for a model with fast agreement dynamics, one might thus consider a variant of a Naming Game with rules modified to induce this active-absorbing phase transition. 

In the present paper, we examine  a  modified Naming Game, which (for $d=2$) undergoes a phase transition between active and absorbing phases. At the transition point, the agreement dynamics proceeds via percolative spreading, which is much faster ($\alpha\sim 1.62$) than the ordinary coarsening as it blocks the formation of stripe structures that do form and slow down the dynamics in an  ordinary Naming Game. 

\section{Model}
Our model is a simple modification of the so-called minimal Naming Game \cite{BARONCHELLI2006,nolfi}, such that agents having more than one word are preferred to those having a single word. In particular, we consider a population of~$N$ agents, placed on sites of the $d$-dimensional Cartesian lattice of linear size $L$ ($N=L^d$), which try to bootstrap a common name for a given object. Each agent has its own  inventory, which is a dynamically modified list of words (empty at the beginning). The following act of communication of agents constitutes an elementary step of the dynamics:
\begin{itemize} 
\item Two agents are picked - one of them is a speaker and the other is a hearer. 
\item The speaker selects randomly a word from its inventory (or generates a new one when it is empty) and transmits  it to the hearer. 
\item If the hearer has the transmitted word in its inventory, the interaction is a success and both players maintain in their inventories only the transmitted word. 
\item If the hearer does not have the transmitted word in its inventory, the interaction is a failure and the hearer updates its inventory by adding to it the transmitted word. 
\end{itemize}
To select a speaker we use the roulette-wheel selection. In this method probability to select a given agent is proportional to its weight. There are various implementations of this method but the one that we used has $O(1)$ computational complexity and  is thus particularly suitable for large $N$ systems \cite{roulette}. The weights of agents in our model are assigned according to the number of words in agents' inventories. They are  equal to unity  for agents with only one word and to~$p$ ($p\geq 1$) for those with more than one word. In other words, $p$~introduces a preference for selecting multi-word agents rather than single-word ones. Once the speaker is selected, the hearer is chosen randomly as one of its nearest neighbours. For the preference $p=1$, our model is, of course, equivalent to the minimal Naming Game.  It is probably difficult to argue that with $p>1$ Naming Game provides a better description of a population of interacting humans trying to bootstrap a common vocabulary (but perhaps such a modification could be implemented in artificial communicating agents). We consider our model rather as a simple modification of a Naming Game that leads to an interesting new behaviour and is thus worth an interest.

To examine our model, we used Monte Carlo simulations. 
To analyse time-dependent characteristics, we define a unit of time as $N$ acts of communication.
The discussion of our results is based mainly on the time-dependent fraction~$x(t)$ of the number of agents having more than one word and the time $\tau$ needed for a system to reach a single-language state. The numerical data presented below are usually averages over many ($10^2\sim10^4$) independent runs; periodic boundary conditions were used for the simulated systems.

\section{Numerical results}
\begin{figure}
\includegraphics[width=\columnwidth]{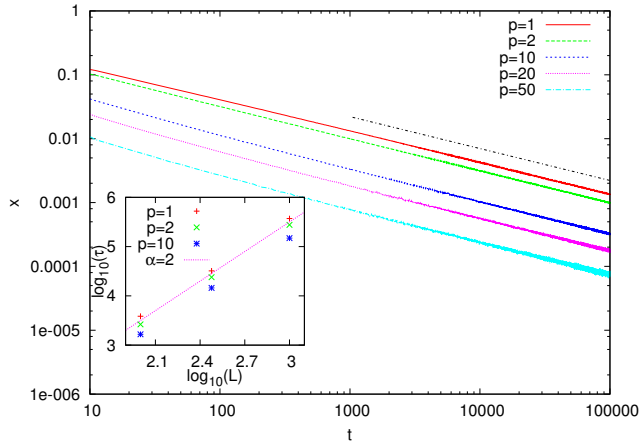}  \vspace{-0.2cm} 
\caption{(Color online) The time dependence of the density of multi-word agents $x(t)$ for a one-dimensional model ($L=10^5$). The dot-dashed straight line corresponds to $x(t)\sim t^{-1/2}$ decay and our numerical data for all examined values of $p$ seem to follow this decay. The inset shows the size dependence of the average time $\tau$ needed for the entire system to reach a single-language state.   
\label{timed1}}
\vspace{-0.2cm}
\end{figure}

Simulations show that in the one-dimensional case our model has the asymptotic decay of $x(t)$ consistent with the decay $t^{-1/2}$ for any $p$ (Fig.~\ref{timed1}). This indicates that qualitatively it behaves as the minimal Naming Game ($p=1$). Indeed,  at late stages of its evolution the model is mainly composed of large single-language domains with agents with more than one word placed typically at their boundaries. It means that the characteristic length of domains~$l$ scales as $l\sim x(t)^{-1}\sim t^{1/2}$, which is a well-known feature of nonconservative coarsening \cite{bray}.  

For $p=1,~2$, and 10, we have also calculated the characteristic time $\tau$ needed to reach a single-language state in the entire system and our numerical data (inset in Fig.~\ref{timed1}) obey the scaling $\tau\sim L^2$, in agreement with Eq.~(\ref{barontau}).

\begin{figure}
\includegraphics[width=\columnwidth]{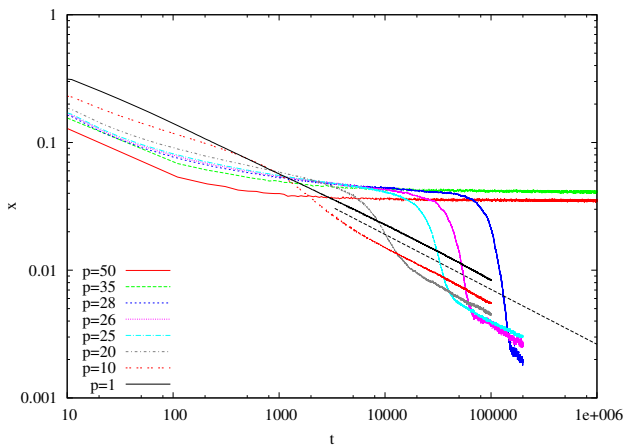}  \vspace{-0.2cm} 
\caption{(Color online) The time dependence of the density of multi-word agents $x(t)$ for the $d=2$ model ($L=1000$). The dashed black line corresponds to $x(t)\sim t^{-0.43}$ decay and numerical data for $p<28$  follow this decay asymptotically. For $p>28$, an asymptotic density $x(t)$ remains positive.
\label{timed2}}
\vspace{-0.2cm}
\end{figure}

The two-dimensional case, however, is more intriguing. For the ordinary minimal Namig Game ($p=1$), we observe that $x(t)$ decays asymptotically as $t^{-0.43}$ (Fig.~\ref{timed2}). Such a decay is obtained from the least-square fit of data from the last decade ($10^4<t<10^5$). Asymptotic decay for $p=10$, 20, 25, and 26 also seems to follow such a decay. By repeating qualitative arguments of the $d=1$ case, we obtain that such a decay implies that the characteristic length $l$ grows as $t^{0.43}$. The increase is thus slower than the expected $l\sim t^{1/2}$, but the accuracy of simulations seems to exclude that on a longer time scale or for a larger $L$, the decay of $x(t)$ will turn into $t^{-1/2}$. Morever, this decay is seen most likely for any $p<28$ and thus it is to some extent universal. 
(A similar universality holds in the Ising model or some other systems quenched below the critical temperature, where, e.g., the characteristic lengthscale increases as a power-law with the exponent independent of the temperature of the quench \cite{bray}.)
Much different behaviour is observed for $p>28$, where simulations show that $x(t)$ does not decay to zero but asymptotically remains positive. It means that there is a positive fraction of agents with more than one word, which indicates that the system does not evolve toward a single-language (absorbing) state but remains in a multi-language (active) state. 

The nature of the phase transition the model undergoes  around $p=28$ between active and absorbing states is not entirely clear. Indeed, it is difficult to distinguish whether at the transition point (Fig.~\ref{timed2}) the density $x(t)$ decays continuously (and most likely slower than a power law) to zero (as in continuous phase transitions) or develops a discontinuous jump (as in discontinuous transitions). Continuous phase transitions are accompanied by critical points, where the density of active sites typically have a power-law decay \cite{haye}.
However, some models with absorbing states and continuous transitions, as for example voter-like models \cite{frach}, are characterized by a much slower (i.e., logarithmic) decay of active sites (which is an analog of $x(t)$ in our model). Such slow decay of the density of active sites was also observed in the two-dimensional Potts model with absorbing states \cite {lipdroz} and, in our opinion, a similar behaviour might be seen in the model examined in the present paper. However, an unambiguous identification of the nature of the phase transition that takes place at $p=28$ would require considerable numerical efforts and is left for the future. Let us also notice that the absorbing state is infinitely degenerate (since any language might be selected as absorbing). For such models, various critical points and phase transitions were reported, including discontinuous ones \cite{haye}. 

\begin{figure}
\includegraphics[width=\columnwidth]{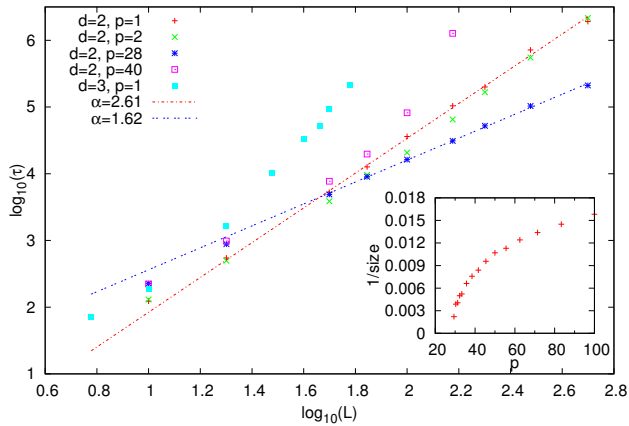}  \vspace{-0.2cm} 
\caption{(Color online) The size dependence of the characteristic time $\tau$ (log-log scale). The inset shows the inverse of the average cluster size of the dominant language as a function of $p$.  
\label{tau}}
\vspace{-0.2cm}
\end{figure}

A slower than $t^{-1/2}$ decay of $x(t)$ indicates that the dynamics for $d=2$ and $p<28$ is slower than the expected coarsening dynamics. A similar conclusion follows from the analysis of the size dependence of the characteristic time $\tau$ (Fig.~\ref{tau}). Indeed, an estimation of $\alpha$ from the asymptotic increase of $\tau$ both  for $p=1$ and 2 yields $\alpha\sim 2.61$, which is considerably larger  than the predicted value of~2~\cite{baron2006}. Some Monte Carlo simulations were made supporting $\alpha=2$  \cite{baron2006}, but they were made for  $L\leq 100$, which is much smaller  than in our simulations ($L\leq 500$). Inside an active phase ($p>28$) the system in principle can also reach the single-language state but the time needed for this process most likely increases exponentially with the system size (Fig.~\ref{tau}). A very interesting behaviour is seen, however, at the transition point $p=28$, where our data clearly indicate a much slower increase of $\tau$ with $\alpha\sim 1.62$.

\begin{figure}
\includegraphics[width=7cm]{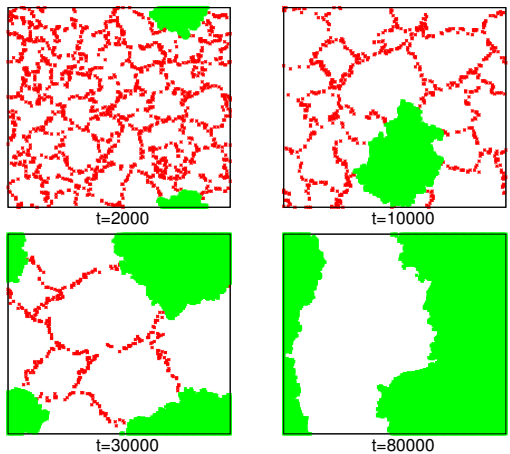}  \vspace{-0.2cm} 
\caption{(Color online) Sample configurations of the model for $p=1$ (an ordinary Naming Game) and $L=200$. Agents with the dominant word in their inventories are shown in green (note periodic boundary conditions in both directions). Agents with more than one word (each different from the dominant one) are shown in red.  Only in a certain fraction of runs such stripe structures are formed. Configurations with several stripes, and several languages might also form. 
\label{confi999}}
\vspace{-0.2cm}
\end{figure}

To trace the origin of a slow dynamics (for $p<28$), we examined snapshot configurations of the model.  We noticed that runs for which the time needed to reach a single-language  state was large proceeed via stripe structures (Fig.~\ref{confi999}). Such stripes are on average zero-curvature configurations, which results in a much different dynamics. They form in some other models as well, as for example in low-temperature Ising models, where they also slow down the dynamics \cite{lip1999,olejarz-prl}. Assuming that the diffusion constant of each interface of size $L$ scales as $1/L$ and that  they are initially separated by a distance of the order of system size $L$, it was argued \cite{lip1999} that the characteristic lifetime of such structures should scale as $L^3$. It might suggest that the asymptotic value of $\alpha$ is 3 and in our simulations  we have not reached a sufficiently large $L$ yet. On the other hand, dynamically generated stripes will be of various thickness (some of them might be quite thin) and in some runs the system will (quickly) evolve without formation of stripes. Thus we cannot exclude that our estimation  $\alpha\sim 2.61$ is actually quite accurate.

\begin{figure}
\includegraphics[width=7cm]{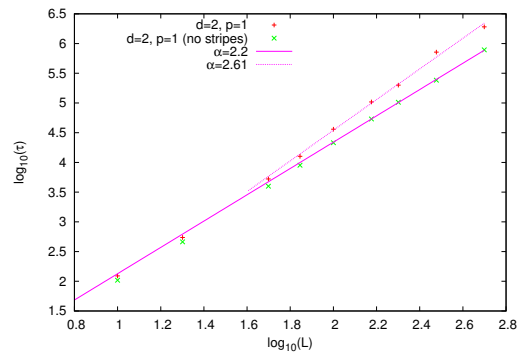}  \vspace{-0.2cm} 
\caption{(Color online) The size dependence of the characteristic time $\tau$ (log-log scale) for $d=2$ and $p=1$. 'no stripes' denotes averaging only over runs that evolve to the single-language state without stripe configurations. The 'no stripes' data can be fitted with $\alpha=2.2$. For comparison we also plotted data where stripe configurations were not discarded.
\label{tau-stripless}}
\vspace{-0.2cm}
\end{figure}

\begin{figure}
\includegraphics[width=4cm]{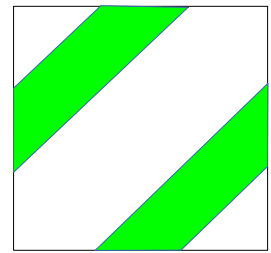}  \vspace{-0.2cm} 
\caption{(Color online) An example of a skewed stripe (note periodic boundary conditions in both directions).
\label{skew-stripes-a}}

\vspace{-0.2cm}
\end{figure}

To show that stripes are indeed responsible for the slow dynamics, we present the size dependence of $\tau$ for $p=1$, where averaging is made only over those runs that did not generate stripes during their evolution to the single-language configuration (Fig.~\ref{tau-stripless}).  To detect stripes in a given run, we were checking whether there existed a vertical or a horizontal row of agents which had only one and the same word in their inventories. Runs with such rows were discarded and instead new runs were generated. Numerical results show that for stripeless configurations the dynamics is indeed faster (Fig.~\ref{tau-stripless}). From the size dependence, we estimate $\alpha\approx 2.2$, which is substantially lower than $2.61$ but still larger than the expected  $\alpha=2$.
Let us notice, however, that our method of detecting stripes is not rigorous since it misses stripes that are skewed (Fig.~\ref{skew-stripes-a}). Such stripes are longer (their length depends on orientation and the winding number) and most likely they have a smaller probability of being created. Nevertheless, such configurations might to some extent affect the dynamics increasing $\alpha$ above 2. Moreover, some bending of numerical data (Fig.~\ref{tau-stripless}) suggests that for larger systems the value might be slightly lower than the estimate $\alpha=2.2$. Leaving for the future a more detailed analysis of the dynamics, we conclude that the existence of stripes (Fig.~\ref{confi999}) along with the fact that their removal substantially reduces $\alpha$ show that stripes are indeed an important factor slowing down the dynamics.

\begin{figure}
\includegraphics[width=7cm]{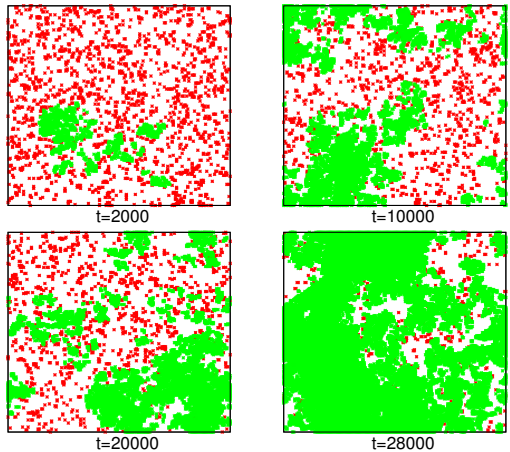}  \vspace{-0.2cm} 
\caption{(Color online) Sample configurations of the model for $p=28$ (transition point) and $L=200$. Agents with the dominant word in their inventories are shown in green. Agents with more than one word (each different from the dominant one) are shown in red. 
\label{config036}}
\vspace{-0.3cm}
\end{figure}

Why is the dynamics at the transition point $p=28$ so fast? Visual inspection of snapshot configurations revealed that they  are much different now (Fig.~\ref{config036}) and prompted us to suggest at least a partial explanation, which refers to percolative properties. First, let us notice that at intermediate stages languages are not restricted to separate domains (as in Fig.~\ref{confi999}) but are dispersed all over the system (in Fig.~\ref{confi999} and Fig.~\ref{config036} only agents using the dominant language are plotted, but the distribution of users of other languages is similar). During the evolution of the system, some languages die out, which increases the concentration of the remaining ones. At a certain moment, the concentration of some language becomes sufficiently large and it forms (effectively) an infinite cluster percolating over the system. If our percolative scenario is correct then the lack of stripes might be related with some percolation theorems about the uniqueness of a percolating cluster in $d=2$ systems \cite{aizenman}.  Formation of such a percolating cluster would break the symmetry (between languages) and preselect the language with which the system will end up. Let us notice, however, that the existence and uniqueness of percolating clusters is an issue, which requires considerable mathematical subtleties. In particular, in $d=2$ systems at the percolation threshold, the unique infinite cluster 'coexists' with infinitely many spanning clusters \cite{aizenman1}.

To get an additional support for the proposed percolative spreading, we calculated the average size of clusters of agents using the dominant language. In the multi-language phase ($p>28$), this quantity increases and possibly even diverges upon approaching the transition point (inset in Fig.~\ref{tau}). This indicates that percolative properties do play an important role at the transition point.

The percolative spreading suggests why stripes do not form and why $\alpha$ should not be greater than~2. It still remains to be understood, however, what makes the dynamics so fast at $p=28$ and why $\alpha$ is considerably smaller than~2. Let us notice that similarly fast coarsening dynamics is known to characterize some quite different physical systems like  two-dimensional binary liquids in the so-called inertial hydrodynamic regime. For such systems numerical simulations \cite{alexander} and some theoretical arguments  \cite{furukawa} suggest that the characteristic length scale increases as $l \sim t^{2/3}$. Naively inverting such a scaling, we obtain that the characteristic time should scale as $\tau \sim L^{3/2}$, which is quite close to the scaling $\tau \sim L^{1.62}$ observed in our model at the transition point $p=28$. It would be certainly interesting to examine whether the above similarity is only a coincidence or it indicates certain relation between these apparently  different dynamics. Let us remark that in the inertial hydrodynamic regime the Reynolds number is large and the system is turbulent, resembling perhaps the structure of our model at the critical point (Fig.~\ref{config036}).

Finally, we comment on the $d=3$ Naming Game. Let us notice that stripe-like structures form also during the coarsening of a low-temperature $d=3$ Ising model \cite{lip1999} and perhaps play a more important role. This is because in the $d=3$ case the stripes are actually two-dimensional membranes, fluctuations and diffusion of which are much slower than of one-dimensional interfaces. As a result, their lifetime might increase even faster than power-law, in agreement with Monte Carlo simulations  of the $d=3$ Ising model \cite{lip1999}. We expect that similar structures form also in the Naming Game. Indeed, our simulations for the ordinary minimal Naming Game show that the characteristic time $\tau$ is likely to increase faster than power-law (data in Fig.~\ref{tau} exhibit some bending, which seems to exclude a power-law fit), which would invalidate Eq.~(\ref{barontau}) in this case. It would be also interesting to examine the three-dimensional version with preference $p$ and check for the existence of a similar phase transition. Let us notice, however, that in the three-dimensional space there might be enough space for more than one infinite cluster and the percolative mechanism might not be such effective in this case.

\section{Conclusions}
In conclusion, we have shown that the agreement dynamics of the Naming Game for $d\geq 2$ is slower than expected, primarily due to the formation of stripes. The existence of absorbing states in the dynamics of the Naming Game suggests a similarity with the zero-temperature nonconservative Ising model \cite{dallasta-phd}, the dynamics of which is also known to generate stripes. However, in the Ising model with zero-temperature dynamics, the stripes have an infinite lifetime. On the other hand, in the Naming Game model, stripes have a finite (albeit sometimes quite large) lifetime and in this regard it resembles the dynamics of the Ising model at low (but positive) temperature. 
However, the dynamics of the Naming Game seems to be richer than a combination of zero- and low-temperature dynamics of Ising model.
Indeed, in the $d=2$ Ising model, the characteristic length scale $l$ is not affected by the  formation of stripes and increases in time as $t^{1/2}$ \cite{lip1999} while in the Naming Game, presumably due to the formation of stripes, we reported a slower increase $l\sim t^{0.43}$. 

We also introduced an extended version of Naming Game that shows a rich and perhaps novel behaviour with intriguing links to some other statistical mechanics problems. Further studies hopefully will clarify the role of percolation transition especially in the context of the $d=3$ version. Possible relation with coarsening dynamics of binary liquids is also worth furhter studies, eventhough at present our arguments with this respect are very speculative and based mainly on some similarities of certain exponents. Finally, let us notice that Naming Game shares some similarities also with a voter model or with an absoring-state Potts model \cite{lipdroz}, as for example symmetric absorbing states and nonconservative dynamics, and our results might be relevant in some wider contexts.

Acknowledgments: D.L. is supported by NCN grant 2011/01/B/HS2/01293 and A.L. is supported by NCN grant 2013/09/B/ST6/02277.

\end {document}